\documentclass[11pt,english]{revtex4}
\usepackage[T1]{fontenc}
\usepackage[latin9]{inputenc}
\usepackage{amsmath}
\usepackage{amssymb}

\makeatletter

\providecommand{\tabularnewline}{\\}

\makeatletter

\makeatletter

\usepackage{mathptmx}

\usepackage{textcomp}

\makeatletter

\usepackage{epsfig}

\usepackage{psfrag}

\makeatother

\makeatother

\makeatother

\usepackage{babel}
\makeatother
\begin{document}

\title{Big-Bang Nucleosynthesis and WIMP dark matter in modified gravity}

\author{Jin U Kang$^{1,2}$ and Grigoris Panotopoulos$^{1}$}

\affiliation{$^{1}$Arnold-Sommerfeld-Center, Department für Physik, Ludwig-Maximilians-Universität
München, Theresienstr. 37, D-80333, Munich, Germany\\
 $^{2}$Department of Physics, Kim Il Sung University, Pyongyang,
Democratic People's Republic of Korea}

\begin{abstract}
In the present work the primordial Big-Bang Nucleosynthesis (BBN)
and weakly interacting massive particle (WIMP) dark matter are discussed
in a certain class of modified gravitational theories, namely $f(R) \sim R^n$ gravity. The new gravitational model
is characterized by a single parameter $n$.
First we determine the conditions under which the theoretical predictions for the $^{4}$He
abundance are in agreement with the observations. More precisely, during BBN the physics is known and all 
the parameters are known. The only free parameter to be constrained is the power $n$ related to the new 
gravitational model. After that, for cold dark matter we use the value of $n$ determined from the BBN considerations
and determine how the mass of the dark matter particle is related to the annihilation cross section in order for the cold dark matter constraint to be satisfied.
\end{abstract}

\maketitle
\newpage{}

\section{Introduction}

There is accumulated evidence both from astrophysics and cosmology that about 1/4 of the energy budget of the universe consists of so called dark matter, namely a component which is non-relativistic and does not feel the electromagnetic nor the strong interaction. For a review on dark matter see e.g.~\cite{Munoz:2003gx}. Although the list of possible dark matter candidates is long (for a nice list see e.g.~\cite{Taoso:2007qk}), it is fair to say that the most popular dark matter particle is the LSP in supersymmetric models with R-parity conservation~\cite{Feng:2003zu}. The superpartners that have the right properties for playing the role of cold dark matter in the universe are the axino, the gravitino and the lightest neutralino. By far the most discussed case in the literature is the case of the neutralino (see the classical review~\cite{Jungman:1995df}), probably because of the prospects of possible detection. On the other hand, primordial Big-Bang nucleosynthesis (BBN) is one of the cornerstones of modern cosmology. In the old days, BBN together with Hubble's law and CMB supported and strengthened  the Hot Big-Bang idea. Nowadays, BBN can be used to test and constrain possible new physics beyond the standard model. The new physics may be either due to exotic particles predicted by particle physics model or due to a new expansion law for the universe predicted by a new gravitational model. For a recent review on BBN see e.g.~\cite{Iocco:2008va}. In the present work we shall be interested in a class of new gravitational models of the form $f(R)\sim R^{n}$, where the power $n$ is the only parameter that characterizes this class of models. Although in the literature the authors usually discuss this kind of modified gravitational models in the late times universe (see e.g.~\cite{Capozziello:2004us, Capozziello:2003gx}), here we wish to discuss this class of models in the early universe. In~\cite{Lambiase:2006dq} the authors were interested in the baryon asymmetry in the framework of gravitational baryogenesis proposed a few years
ago~\cite{Davoudiasl:2004gf}.  

In this Letter we wish to study this class of gravity models in two respects,
namely primordial Big-Bang nucleosynthesis (BBN) and WIMP dark
matter. Our investigation will allow us to first derive the allowed range for the
power $n$, and to see how different this class of models can be compared to general relativity. Then for these values of $n$ we determine how the WIMP mass has to be related to its annihilation cross section so that the cold dark matter constraint is satisfied. As a matter of fact, already in \cite{Lambiase:2006dq}, the authors have mentioned that obtaining the right baryon asymmetry in agreement with BBN requires a value of $n$ close to unity. Their discussion was based on the argument that the temperature relevant for BBN should be within the range $0.1-100$~MeV. Here, however, we perform a more accurate investigation by actually computing the cosmological helium abundance employing the semi-analytical method introduced in~\cite{key1}. We remark that one can use numerical codes~\cite{Wagoner:1966pv} for a proper treatment of BBN and accurate computation of the light nuclei abundances. However, the final density of $^4He$ is very weakly sensitive to the whole nuclear network~\cite{Iocco:2008va}. Therefore, in the present investigation we shall employ the semi-analytical treatment of~\cite{key1}, computing the Helium abundance to a very good approximation avoiding sophisticated computer softwares. See also~\cite{Fabris}
for a recent example of a published work in which the same semi-analytical method was used to constrain the higher dimensional Planck mass in a brane model.
Our results show that BBN requires the models considered in the present work to be only slightly different from the usual Einstein's general
relativity, whereas dark matter consideration alone does not seem to constrain this class of new gravity theories due to the degeneracy in paramater space of the underlying particle physics models. However, from the BBN consideration we give a precise range for the allowed values of the power $n$  and confirm the result of~\cite{Zakharov:2006uq} using a different approach based on physics of the early universe. Finally, we remark at this point that according to our findings, certain scenarios that require a value of $n$ considerably different than one cannot work. Furthermore, the models that satisfy our constraints do not lead to the late cosmic acceleration.

Our work is organized as follows. The article consists of five sections,
of which this introduction is the first. The modified gravitational
model is described in the next section. The analysis based on BBN
is discussed in section 3, while the investigation based on WIMP
dark matter is presented in section 4. Finally we conclude in the
last section.

\section{The modified gravitational model}

Here we shall present the model of $f(R)$ gravity that will be discussed
in this paper, and we shall summarize the basic formulas following~\cite{Lambiase:2006dq}.
The model is described by the action \begin{equation}
S=\frac{1}{2\kappa^{2}}\int d^{4}x\sqrt{-g}f(R)+S_{m}[g_{\mu\nu},\phi_{m}],\label{eq:general action}\end{equation}
 where $G$ is Newton's constant, $\kappa^{2}=8\pi G$, and $S_{m}$
is the action of the matter field, $\phi_{m}$. Varying this action
with respect to the metric we obtain the field equations for gravity,
which generalize the usual Einstein's equations, \begin{equation}
f'R_{\mu\nu}-\frac{1}{2}fg_{\mu\nu}-\nabla_{\mu}\nabla_{\nu}f'+g_{\mu\nu}\Box f'=\kappa^{2}T_{\mu\nu},\label{eq:field eq}\end{equation}
 where $T_{\mu\nu}$ is the energy-momentum tensor for the matter,
and a prime denotes differentiation with respect to $R$. For the
gravity part we consider the spatially flat Robertson-Walker (RW)
line element \begin{equation}
ds^{2}=dt^{2}-a(t)^{2}(dx^{2}+dy^{2}+dz^{2}),\label{eq:FRW}\end{equation}
 while for the matter part we consider a cosmological fluid characterized
by a time-dependent energy density $\rho(t)$ and pressure $p(t)$
\begin{equation}
T_{\nu}^{\mu}=diag(\rho,-p,-p,-p).\label{eq:EMT}\end{equation}
 The $0-0$ component of \eqref{eq:field eq} gives \begin{equation}
-3\frac{\ddot{a}}{a}f'-\frac{1}{2}f+3\frac{\dot{a}}{a}f''\dot{R}=\kappa^{2}\rho,\label{eq:0-0}\end{equation}
 while the $i-i$ components give \begin{equation}
\left(\frac{\ddot{a}}{a}+2\frac{\dot{a}^{2}}{a^{2}}\right)f'+\frac{1}{2}f-2\frac{\dot{a}}{a}f''\dot{R}-f'''\dot{R}^{2}-f''\ddot{R}=\kappa^{2}p.\label{eq:1-1}\end{equation}
 Here a dot denotes differentiation with respect to the cosmic time
$t$. In addition to the above cosmological equations, we have as
usual the energy conservation law \begin{equation}
\dot{\rho}+3\frac{\dot{a}}{a}(\rho+p)=0.\label{eq:energy conservation}\end{equation}
 We restrict ourselves to the models of the form \begin{equation}
f(R)=\left(\frac{R}{A}\right)^{n},\label{eq:f(R)}\end{equation}
 where $A$ is a constant, $A\sim M_{p}^{2-2/n}$, with $M_{p}=1.22\times10^{19}$~GeV
the Planck mass. The power $n$ is the unique parameter of this class
of modified gravitational models, and $n=1$ corresponds to the usual
Einstein's theory. Since we are interested in the physics of the early
universe, we consider the radiation dominated era in which $p=\rho/3$,
and $\rho\sim a^{-4}$. Searching for a power law solution for the
scale factor, $a(t)\sim t^{\alpha}$, the cosmological equations determine
the unknown power $\alpha$ in terms of $n$ as follows: \begin{equation}
\alpha=\frac{n}{2}.\label{eq:n and alpha}\end{equation}
 Notice that when $n=1$ we recover the known result $a(t)\sim t^{1/2}$
for the radiation era in the usual Einstein's general relativity.
Then using the expression for the energy density \begin{equation}
\rho=\frac{\pi^{2}}{30}g_{*}T^{4},\label{eq:rho}\end{equation}
 one obtains the relation between time and temperature \begin{equation}
T=\left(\frac{15}{4\pi^{3}g_{*}}\right)^{1/4}g_{\alpha}^{1/4}\frac{M_{p}^{1/2}}{t^{\alpha}A^{\alpha/2}},\label{eq:T and t}\end{equation}
 where $g_{*}$ counts the relativistic degrees of freedom for energy
density, and \begin{equation}
g_{\alpha}\equiv6^{2\alpha}\alpha^{2\alpha}\frac{-10\alpha^{2}+8\alpha-1}{2(1-2\alpha)^{1-2\alpha}}.\label{eq:g_alpha}\end{equation}
 Since the quantity $g_{\alpha}$ must be positive, the allowed range
of $\alpha$ is $0.155\lesssim\alpha\lesssim1/2$, and $\alpha=1/2$
corresponds to Einstein's theory. Finally the Hubble parameter is
given by \begin{equation}
H(T)=\frac{\alpha A^{\frac{1}{2}}}{g_{\alpha}^{\frac{1}{4\alpha}}M_{p}^{\frac{1}{2\alpha}}}\left(\frac{4\pi^{3}g_{*}}{15}\right)^{\frac{1}{4\alpha}}T^{\frac{1}{\alpha}}\label{eq:Hubble parameter_T}\end{equation}
 which generalizes the usual formula $H(T)\sim T^{2}$, valid in the
standard cosmology of Einstein's general relativity.

\section{Big-Bang Nucleosynthesis}

In this section we briefly review the sequence of basic events leading
to the synthesis of primordial Helium during the early stages of the
expansion of the universe in the standard cosmology based on Einstein's
general relativity, following~\cite{key1}. Then we shall present the corresponding discussion and
our results for the modified gravity case. Since all the parameters are fixed and the power $n$ is the only free parameter of the model, our discussion will allows us to determine the allowed range for $n$. 

When the rates of the weak interactions keeping baryons in chemical equilibrium with leptons become comparable to the Hubble parameter, the neutron fraction $X=n_n/(n_n+n_p)$ is frozen at some value $X(T\simeq0)$ to be determined below, where $n_n$ and $n_p$ are neutron and proton number density respectively. Once the temperature has fallen below about $1/25$ of the Deuterium binding energy, the Deuterium bottleneck opens up, and nearly all of the original neutrons
present at the decoupling time are captured in $^{4}He$. Taking into account the neutron decay, the final helium mass fraction is given by 
\begin{equation} \label{Y4}
Y_{4} \simeq 2exp(-t_{c}/\tau)X(T\simeq0) 
\end{equation} 
where $\tau=885.7\pm0.8$~sec~\cite{key2} is neutron's lifetime,
and $t_{c}\sim3$min is the capture time at which neutrons are captured
into Deuterium.

Now we discuss how to compute $X(T\simeq0)$ and $t_{c}$. To this end, we employ a semi-analytical method
(see e.g.~\cite{key1}, \cite{Muk}) which
is sufficiently accurate and very useful, since the
physics is very transparent and the dependence of the abundances
on input parameters can be clearly worked out without sophisticated computer softwares. Indeed, the abundance of $^4He$ is very weakly sensitive to the whole nuclear network~\cite{Iocco:2008va}, and therefore a sufficiently accurate result can be obtained without using computer codes. Our semi-analytical approach relies on the work of \cite{key1}, and we shall not present here in detail
all relevant formulas, as they are quite involved. To compute $X(T\simeq0)$ we need to integrate the following
rate equation
\begin{equation}
\frac{dX(t)}{dt}=\lambda_{pn}(t)(1-X(t))-\lambda_{np}X(t).\label{eq:rate equation}\end{equation}
 Here we denote by $\lambda_{pn}$ the rate for the weak processes
to convert protons into neutrons and by $\lambda_{np}$ the rate for
the reverse processes that convert neutrons into protons. These rates
are time dependent because of their temperature dependence. The rate
$\lambda_{np}$ is the sum of the rates of three processes \begin{equation}
\lambda_{np}=\lambda(\nu+n\to p+e^{-})+\lambda(e^{+}+n\to p+\bar{\nu})+\lambda(n\to p+\bar{\nu}+e^{-}),\label{eq:Lambda_np}\end{equation}
 each of which is computed using standard field-theoretic techniques.
After a few simplifications the rate $\lambda_{np}$ is computed as
follows~\cite{key1}. \begin{equation}
\lambda_{np}(y)=\left(\frac{255}{\tau y^{5}}\right)(12+6y+y^{2}),\label{eq:lambda(y)}\end{equation}
 where $y=\Delta m/T$ with $\Delta m=m_{n}-m_{p}=1.29$~MeV being the neutron-proton mass
difference. The detailed balance relation gives
\begin{equation}
\lambda_{pn}(y)=e^{-y}\lambda_{np}(y).\label{eq:detailed balance}\end{equation}
 We now rewrite \eqref{eq:rate equation} in terms of $y$ instead
of time as \begin{equation}
\frac{dX(y)}{dy}=\frac{dt}{dy}\left(\lambda_{pn}(y)(1-X(y))-\lambda_{np}(y)X(y)\right),\label{eq:rate equation for X(y)}\end{equation}
 where $dt/dy$ can be computed using the definition of $y$, $y=\Delta m/T$,
and the fact that $\dot{T}/T=-H$. With the initial condition $X(y=0)=1/2$,
the rate equation for $X(y)$ can be integrated numerically, and from
the graphical solution one can compute $X(T\simeq0)$, which we denote
by $\bar{X}$.

Finally, let us add a few words regarding the capture time. The bottleneck
opens up when the main reaction converting Deuterium into heavier
elements \begin{equation}
D+D\rightarrow T+p\end{equation}
 become efficient. First we introduce \begin{equation}
z=\frac{\epsilon_{D}}{T}, \end{equation}
where  $\epsilon_D=m_p+m_n-m_D=2.23$~MeV is the Deuterium binding energy. The another quantity that is important in estimating capture time is the Deuterium abundance, $X_{D}\equiv n_{D}/(n_{n}+n_{p})$, where $n_D$ is the Deuterium number density.
From the Saha equation $X_D$ is given by \begin{equation}
X_{D}=2.8\times10^{-14}\eta_{10}T_{MeV}^{3/2}exp(z)X_{p}X,\end{equation}
where $X_p=n_p/(n_n+n_p)$ is the proton fraction.
In the above formula $T_{MeV}$ is the temperature in $\textrm{MeV}$
units, and we have parametrized the baryon-to-photon ratio by \begin{equation}
\eta_{10}\equiv10^{10}\times\frac{n_{b}}{n_{\gamma}},\end{equation}
 where we use the observational value $n_{b}/n_{\gamma}=6.1\times10^{-10}$
from WMAP~\cite{wmap}. The condition that determines the temperature
(or time) at which the Deuterium bottleneck opens up reads as follows(for
more details we refer the reader to \cite{key1}). \begin{equation}
2X_{D}R_{DD}\simeq1,\label{condition}\end{equation}
 where \begin{equation}
R_{DD}=\frac{dt}{dz}<\sigma\upsilon>n_{b}=2.9\times10^{7}z^{-4/3}exp(-1.44z^{1/3})\end{equation}
 and $<\sigma\upsilon>$ is the thermal average of the relevant cross
section times relative velocity, which is a function of $z$. Then
one can solve \eqref{condition} with respect $z$ to get $t_{c}$,
and then from \eqref{Y4} one can finally obtain $Y_{4}$.

We now consider the constraints on $f(R)$ gravity coming from BBN
by computing the Helium mass fraction at the conclusion of the BBN.
To this end, we will follow \cite{key1} for standard cosmology as described above, and
the modifications will be done by adopting the relation \eqref{eq:T and t}
instead of the standard one.

In the modified gravity model, the basic physics governing the details
of primordial nucleosynthesis remains the same, and the only thing
that is different now is the new time-temperature relation \eqref{eq:T and t},
from which one obtains \begin{equation}
t=\left(\frac{15}{4\pi^{3}10.75}\right)^{1/4\alpha}\frac{M_{p}^{1/2\alpha}}{A^{1/2}}\left(\frac{y}{\Delta m}\right)^{1/\alpha}g_{\alpha}^{1/4\alpha},\label{eq:time_y}\end{equation}
 \begin{equation}
t=\left(\frac{15}{4\pi^{3}3.37}\right)^{1/4\alpha}\frac{M_{p}^{1/2\alpha}}{A^{1/2}}\left(\frac{z}{\varepsilon_{D}}\right)^{1/\alpha}g_{\alpha}^{1/4\alpha}.\label{eq:time_z}\end{equation}
 In \eqref{eq:time_y}-\eqref{eq:time_z} we have taken into account
the appropriate value for $g_{*}$ at the relevant temperature. Finally,
the $^{4}He$ mass fraction is still given by \eqref{Y4},
but now both freeze-out abundance $X(T\simeq0)=\bar{X}$ and capture
time $t_{c}$ are modified due to the new time-temperature relation.

\begin{table}[h]
\begin{tabular}{|c|c|c|c|}
\hline
$\delta=1-n$ &
$\bar{X}$ &
$t_{c}$(sec) &
$Y_{4}$\tabularnewline
\hline
\hline
$0$ &
0.1529 &
176.76 &
0.2504\tabularnewline
\hline
$10^{-5}$ &
0.1528 &
176.94 &
0.2503\tabularnewline
\hline
$10^{-4.5}$ &
0.1526 &
177.34 &
0.2499\tabularnewline
\hline
$10^{-4}$ &
0.1521 &
178.59 &
0.2486\tabularnewline
\hline
$10^{-3.9}$ &
0.1519 &
179.06 &
0.2482\tabularnewline
\hline
$10^{-3.8}$ &
0.1516 &
179.66 &
0.2476\tabularnewline
\hline
$10^{-3.7}$ &
0.1513 &
180.42 &
0.2468\tabularnewline
\hline
$10^{-3.6}$ &
0.1509 &
181.37 &
0.2459\tabularnewline
\hline
$10^{-3.5}$ &
0.1504 &
182.58 &
0.2448\tabularnewline
\hline
$10^{-3.4}$ &
0.1497 &
184.12 &
0.2433\tabularnewline
\hline
$10^{-3.3}$ &
0.1489 &
186.07 &
0.2414\tabularnewline
\hline
$10^{-3.2}$ &
0.1479 &
188.54 &
0.2391\tabularnewline
\hline
$10^{-3.1}$ &
0.1466 &
191.71 &
0.2362\tabularnewline
\hline
$10^{-3}$ &
0.1450 &
195.77 &
0.2326\tabularnewline
\hline
$10^{-2.9}$ &
0.1430 &
201.01 &
0.2280\tabularnewline
\hline
\end{tabular}

\caption{Helium abundance, capture time, and freeze-out neutron mass fraction
for several values of $\delta=1-n$. \label{tab:tab1}}
\end{table}

First we integrate \eqref{eq:rate equation for X(y)} with the initial
condition $X(y=0)=1/2$ to obtain $\bar{X}$ (it is enough
to evaluate $\bar{X}$ at $y=15$ because it freezes out). Then we
use the condition (\ref{condition}) to compute $t_{c}$. Note that
now the function $t(T)$ is the one predicted by the new gravitational
model. Finally we compute $Y_{4}$ from \eqref{Y4}
for several values of $\alpha$ (or $n$). Our results are shown in
Table \ref{tab:tab1} and Fig.~1. The results show that $Y_{4}$
is very sensitive to $1-n$. The freeze-out abundance decreases, while
the capture time increases, as $1-n$ increases (or $\alpha$ decreases).
These both effects bring about decrease in $Y_{4}$ as $1-n$ increases.
The reason is as follows. The decrease of the freeze-out abundance
is due to the decrease of the freeze-out temperature, which is calculated
by equating the Hubble parameter \eqref{eq:Hubble parameter_T} with
the weak interaction rate $\Gamma\equiv\lambda_{np}(T)+\lambda_{pn}(T)\sim T^{5}$\cite{Kolb:1988aj}.
The increase of the capture time is mainly due to the temperature-time
relation: For fixed temperature, the corresponding time increases
as $\alpha$ decreases.

We have interpolated the computed helium abundances as a function
of $n$ (Fig.~1) and obtained the lower bound of $n$ in comparison
with the observational data. The allowed range for $n$ is \begin{equation}
1-n\lesssim0.00016\,\,\,\,\,\text{\,\,\,\,\,\,\,\,\,\,\,\, or equivalently}\,\,\,\,\,\,\,\,\,\,\,\,\,\,\alpha\gtrsim0.49992\label{eq:BBN constraint}\end{equation}
 according to the observational constraint~\cite{Izotov:2007ed}
 \begin{equation}
Y_{4}=0.2516\pm0.0040.\label{eq:observed constraint on Y4}
\end{equation}
We see that according to our results the power $n$, compared to the value found in~\cite{Lambiase:2006dq}, should be pushed even closer to unity. This implies that the predicted baryon asymmetry within the framework of gravitational baryogenesis using this class of models becomes now too low. Furthermore, several existing scenarios that require a value of $n$ considerably different than unity cannot work.

\section{WIMP dark matter}

Recent cosmological observations~\cite{wmap} have established the
allowed range of the normalized density of cold dark matter in the
universe
\begin{equation}
0.075\lesssim\Omega_{cdm}h^{2}\lesssim0.126.\label{eq:observed relic density}
\end{equation}
In the present section we assume that the role of cold dark matter
in the universe is played by weakly-interacting massive particles
(WIMPs) proposed by physics beyond the standard model. For concreteness
one can think of the lightest neutralino. The discussion to follow is a model-independent one, 
in which we have considered a generic WIMP assuming that its mass is ($100-500$)~GeV,
and that its typical cross section of the relevant processes in which
it participates is not very different than $\sigma\sim\alpha_{em}^{2}/M_{ew}^{2}$, where
$\alpha_{em}=1/137$ is the electromagnetic fine structure constant,
and $M_{ew}\sim100$~GeV is the electroweak scale. The relic density of
the dark matter particle depends on its mass $m$, its annihilation cross section $\sigma_0$, and finally on the power $n$ characterizing the gravitational model. Since BBN has already determined the allowed range for $n$, we fix it to a given value and therefore the cold dark matter constraint $\Omega h^2 \simeq 0.1$ gives a certain relation between the WIMP mass and the annihilation cross section. 

The evolution of the number density $n$ of the dark matter particle
in an expanding universe is determined by solving the Boltzmann equation~\cite{Kolb:1988aj,Feng:2003zu}
\begin{equation}
\dot{n}+3Hn=-<\sigma\upsilon>\:(n^{2}-n_{EQ}^{2}),\end{equation}
 where $H$ is the Hubble parameter, $n_{EQ}$ is the number density
at equilibrium, $\upsilon$ is the relative velocity, and $\sigma$
is the total annihilation cross section. The thermal average of the
total annihilation cross section times the relative velocity $<\sigma\upsilon>$
is given by \begin{equation}
<\sigma\upsilon>=\frac{1}{n_{EQ}^{2}}\int\frac{d^{3}p_{1}}{(2\pi)^{3}}\frac{d^{3}p_{2}}{(2\pi)^{3}}f(E_{1})f(E_{2})\sigma\upsilon,\end{equation}
 where $f(E)$ is the fermion distribution function, $f(E)=1/(1+exp(E/T))$.
Finally the number density at equilibrium is given by
\begin{equation}
n_{EQ}=\int\frac{d^{3}p}{(2\pi)^{3}}f(E).
\end{equation}
 Now it is convenient to introduce new variables, namely dimensionless
quantities
\begin{eqnarray}
x & = & \frac{m}{T},\\
Y & = & \frac{n}{s},
\end{eqnarray}
where $T$ is the temperature and $s$ is the entropy density
\begin{equation}
s=h_{*}\frac{2\pi^{2}}{45}T^{3}
\end{equation}
 with $h_{*}$ being the number of relativistic degrees of freedom
for entropy density. Assuming entropy conservation, the Boltzmann
equation can be written down equivalently as follows
\begin{equation}
\frac{dY}{dx}=-\frac{s}{xH}<\sigma\upsilon>(Y^{2}-Y_{EQ}^{2}).\label{eq:Boltzmann equation _Y}
\end{equation}
 The yield at equilibrium $Y_{EQ}$ for non-relativistic (cold, $x\gg3$)
relics is given by the approximate expression
\begin{equation}
Y_{EQ}\simeq g\frac{45}{2\pi^{4}}\left(\frac{\pi}{8}\right)^{1/2}\frac{x^{3/2}exp(-x)}{h_{*}},
\end{equation}
where $g=2$ is the spin polarizations of the dark matter particle.
In standard cosmology during the radiation dominated era, the Hubble
parameter as a function of the temperature is given by $H(T)=1.67g_{*}^{1/2}T^{2}/M_{p}$.
Parameterizing $<\sigma\upsilon>$ as \begin{equation}
<\sigma\upsilon>=\sigma_{0}x^{-l}\end{equation}
 the Boltzmann equation takes the final compact form \begin{equation}
\frac{dY}{dx}=-\lambda x^{-l-2}(Y^{2}-Y_{EQ}^{2}),\label{eq:dy/dx}\end{equation}
 where $\lambda$ is a constant given by \begin{equation}
\lambda=\left(\frac{x<\sigma\upsilon>s}{H(m)}\right)_{x=1}=0.264(h_{*}/g_{*}^{1/2})M_{p}m\sigma_{0}.\label{eq:lambdast}\end{equation}
 We can obtain an approximate analytical solution of Boltzmann equation
by the following arguments. Initially, for large temperatures the
annihilation rate is larger than the expansion rate of the universe
and the WIMP abundance follows the equilibrium abundance. At
some point $x_{f}$ the annihilation rate becomes comparable to the
expansion rate and the dark matter particle decouples from the thermal bath.
For $x\gg x_{f}$ we can neglect the $Y_{EQ}$ term in the Boltzmann
equation. Then the equation can be easily integrated and the solution
$Y_{\infty}\equiv Y(x=\infty)$ is given by \begin{equation}
Y_{\infty}=\frac{l+1}{\lambda}x_{f}^{l+1},\end{equation}
 where the freeze-out temperature $x_{f}$ is given\cite{Kolb:1988aj}
by \begin{eqnarray}
x_{f} & = & \ln[0.038(l+1)(g/g_{*}^{1/2})M_{p}m\sigma_{0}]\label{eq:x_f}\\
 &  & -\left(l+\frac{1}{2}\right)\ln\{\ln[0.038(l+1)(g/g_{*}^{1/2})M_{p}m\sigma_{0}]\}.\nonumber \end{eqnarray}
 After having integrated the Boltzmann equation for $Y(x)$, then
the relic abundance for the dark matter particle is given by
\begin{equation}
\Omega_{cdm}h^{2}=\frac{mY_{\infty}s(T_{0})h^{2}}{\rho_{cr}},\end{equation}
 where $T_{0}$ is the today's temperature. Here we make use of the
following values:
\begin{eqnarray}
T_{0} & = & 2.73K=2.35\times10^{-13}~GeV\\
h_{*}(T_{0}) & = & 3.91\\
\rho_{cr}/h^{2} & = & 8.1\times10^{-47}~GeV^{4}
\end{eqnarray}
So far we discussed the case of the standard cosmology. Now let us
take into account the modification of gravity. Taking into account
$a\sim t^{\alpha}$ and time-temperature relation \eqref{eq:T and t},
one can express the Hubble parameter as \begin{equation}
H=H_{\alpha}(m)\, x^{-1/\alpha},\label{eq:H_H_alpha}\end{equation}
 where \begin{equation}
H_{\alpha}(m)=\frac{\alpha A^{\frac{1}{2}}}{g_{\alpha}^{\frac{1}{4\alpha}}M_{p}^{\frac{1}{2\alpha}}}\left(\frac{4\pi^{3}g_{*}}{15}\right)^{\frac{1}{4\alpha}}m^{\frac{1}{\alpha}}.\label{eq:H_alpha}\end{equation}
 For $\alpha=1/2,$ this reduces to the usual $H(m)$ parameter for
the standard cosmology~\cite{Kolb:1988aj} \begin{equation}
H_{\alpha=1/2}(m)=1.67g_{*}^{1/2}m^{2}/M_{p}=H(m).\label{eq:H(m)}\end{equation}
 For $x\gtrsim3$ the temperature dependence of the annihilation cross
section is parameterized as \begin{equation}
<\sigma\upsilon>\equiv\sigma_{0}x^{-l},\label{eq:appoximation for sigma_v}\end{equation}
 where $l=0$ corresponds to s-wave annihilation, $l=1$ to p-wave
annihilation, etc. Then the Boltzmann equation for the abundance of
dark matter becomes \begin{equation}
\frac{dY}{dx}=-\tilde{\lambda}x^{-\tilde{l}-2}(Y^{2}-Y_{EQ}^{2}),\label{eq:Boltzmann_Y}\end{equation}
 where \begin{equation}
\tilde{\lambda}=\left(\frac{x<\sigma\upsilon>s}{H_{\alpha}(m)}\right)_{x=1}=\frac{H(m)}{H_{\alpha}(m)}\lambda=0.264(h_{*}/g_{*}^{1/2})M_{p}m\tilde{\sigma}_{0}.\label{eq:lambda}\end{equation}
 Here we introduced new parameter $\tilde{l}$ and $\tilde{\sigma}_{0}$
to clearly show the effect of the modification of gravity on Boltzmann
equation: \begin{equation}
\tilde{l}=l+(2-\frac{1}{\alpha}),\label{eq:power of x in crosssection}\end{equation}
 \begin{equation}
\tilde{\sigma}_{0}=\frac{H(m)}{H_{\alpha}(m)}\sigma_{0}.\label{eq:sigma_0}\end{equation}
It follows that $\tilde{l}=l$ and $\tilde{\sigma}_{0}=\sigma_{0}$
for $\alpha=1/2$. It is clear that the modification coming from $f(R)$
is implemented entirely by the correction on these two parameters.
By comparison to \eqref{eq:dy/dx}-\eqref{eq:lambdast}, one can easily
see that the Boltzmann equation \eqref{eq:Boltzmann_Y} together with
\eqref{eq:lambda} exactly corresponds to one of the standard cosmology
with the averaged product of annihilation cross section and velocity
\begin{equation}
<\sigma\upsilon>=\tilde{\sigma}_{0}x^{-\tilde{l}}.\label{eq:sigma times v}
\end{equation}
Since the Boltzmann equation has exactly the form as in
standard cosmology, one would get the same results as in the standard
case, but with the replacement, $\sigma_{0}\to\tilde{\sigma}_{0}$
and $l\to\tilde{l}_{0}$.

The most important quantity in estimating the relic density is $x_{f}$,
which is the time when $Y$ ceases to track $Y_{EQ}$, or equivalently,
when $Y-Y_{EQ}$ becomes of order $Y_{EQ}$. This quantity is computed
by \eqref{eq:x_f} as
\begin{eqnarray}
x_{f} & = & \ln[0.038(\tilde{l}+1)(g/g_{*}^{1/2})M_{p}m\tilde{\sigma}_{0}]\label{eq:x_f_tilde}\\
 &  & -\left(\tilde{l}+\frac{1}{2}\right)\ln\{\ln[0.038(\tilde{l}+1)(g/g_{*}^{1/2})M_{p}m\tilde{\sigma}_{0}]\}.\nonumber \end{eqnarray}
Then the present yield $Y_{\infty}$ and relic density
$\Omega_{cdm}h^{2}$ are given by
\begin{equation}
Y_{\infty}=\frac{3.79(\tilde{l}+1)x_{f}^{\tilde{l}+1}}{(h_{*}/g_{*}^{1/2})M_{p}m\tilde{\sigma}_{0}},\label{eq:present relic abundance}
\end{equation}
\begin{equation}
\Omega_{cdm}h^{2}=1.07\times10^{9}\frac{(\tilde{l}+1)x_{f}^{\tilde{l}+1}\text{GeV}^{-1}}{(h_{*}/g_{*}^{1/2})M_{p}\tilde{\sigma}_{0}}.\label{eq:relic density}
\end{equation}
Since the cosmic temperature during the period of interest is $T\simeq$
(a few)GeV, we set $h_{*}=g_{*}\simeq100$.
Eq. \eqref{eq:x_f_tilde} - \eqref{eq:relic density} are one of the main results of this section. It is remarkable that they are expressed as an analytical function of $\alpha$ through $\tilde{l}$, $\tilde{\sigma}_{0}$ and $x_{f}$. Thus once $l$, $\sigma_0$  and $m$ together with $\alpha$ are given, the relic density is directly obtained from \eqref{eq:relic density}.
 We will use s-wave approximation
($l=0$), so $\tilde{l}=2-1/\alpha$. In this case $\tilde{l}$ is
negative because $\alpha<1/2$, but since we know from the consideration
in the previous section that BBN allows tiny deviation of $n$ from
$1$ in \eqref{eq:f(R)}, we assume that $\tilde{l}+1>0$ in \eqref{eq:x_f_tilde}-\eqref{eq:relic density}.

We are finally in a position to present our numerical results in figures, showing the relation between the annihilation cross section and WIMP mass. The dark matter abundance is a function of three parameters, namely $n, m, \sigma_0$. If the power $n$ is fixed according to the BBN results, and we impose the cold dark matter constraint $0.075 < \Omega_{cdm} h^2 < 0.126$, it is possible to obtain a certain relation between the WIMP mass $m$ and its annihilation cross section $\sigma_0$. Our results can be shown in the figures 2 and 3 below. In particular, in Fig.~2 we show the annihilation cross section as a function of the WIMP mass for fixed values of $n$ (corresponding either to general relativity or to the new gravitational model for the range determined from BBN), and for the upper limit $\Omega_{cdm} h^2=0.126$. In Fig.~3 we show $\sigma_0$ as a function of $m$ for fixed values of $n$ and for the lower limit $\Omega_{cdm} h^2=0.075$. From these figures one can see what the lower bound (Fig.~2), and upper bound (Fig.~3) of the annihilation cross section should be for a given WIMP mass.

\section{Conclusions}

In the present work we have studied primordial Big-Bang nucleosynthesis
and WIMP dark matter in a class of modified gravitational theories.
This class of gravitational models predict a novel expansion law for
the early universe. For BBN we have
employed a semi-analytical computation in which the basic physics
is quite transparent. For WIMP dark matter we have given a model independent discussion applying the
usual treatment found in standard textbooks or reviews.
Concerning BBN, by comparing the theoretical predictions to the available observational data we were able to put bounds on the unique parameter appearing in this class of modified gravitational theories. We have found that the models considered in the present work are allowed to be only slightly different from the usual Einstein's general relativity. In the dark matter section we have obtained an analytical expression for the cold dark matter abundance as a function of n. After that we fixed the power $n$ according to the BBN results, and we have shown in figures how the annihilation cross section and the WIMP mass are related in order that the cold dark matter constraint is satisfied. The predicted baryon asymmetry within the framework of gravitational baryogenesis using this class of models~\cite{Lambiase:2006dq} is too low for our BBN range of $n$, and therefore this mechanism for baryon asymmetry does not seem to be consistent with BBN constraints. Finally we remark in passing that the models that satisfy our bounds do not lead to the late cosmic acceleration.

\begin{acknowledgments}
We thank the anonymous reviewer for valuable comments and suggestions. J.~U K. is supported by the German Academic Exchange Service (DAAD),
and G.~P. is supported by project ``Particle Cosmology''.
\end{acknowledgments}

\newpage{}

\begin{figure}
\centerline{\epsfig{figure=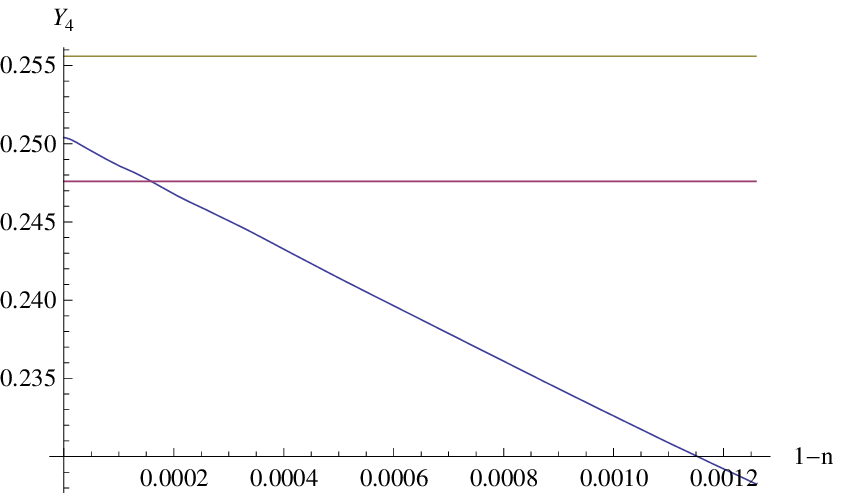,height=8cm,angle=0}}
\caption{Theoretical helium 4 abundance versus $\delta=1-n$. The strip shows
the allowed observational range.}
\end{figure}


\newpage{}

\begin{figure}
\centerline{\epsfig{figure=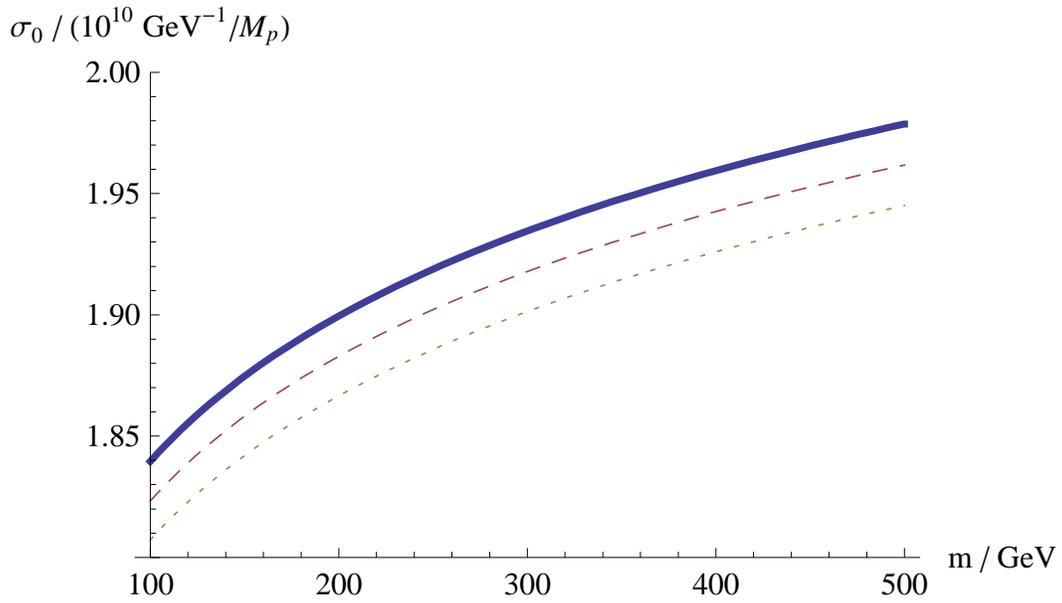,height=8cm,angle=0}}
\caption{Annihilation cross section versus WIMP mass for CDM abundance $\Omega_{cdm} h^2=0.126$. Shown are $n=1$ 
(solid), $n=1-10^{-4}$ (dashed), and $n=1-2 \times 10^{-4}$ (dotted).}
\end{figure}

\begin{figure}
\centerline{\epsfig{figure=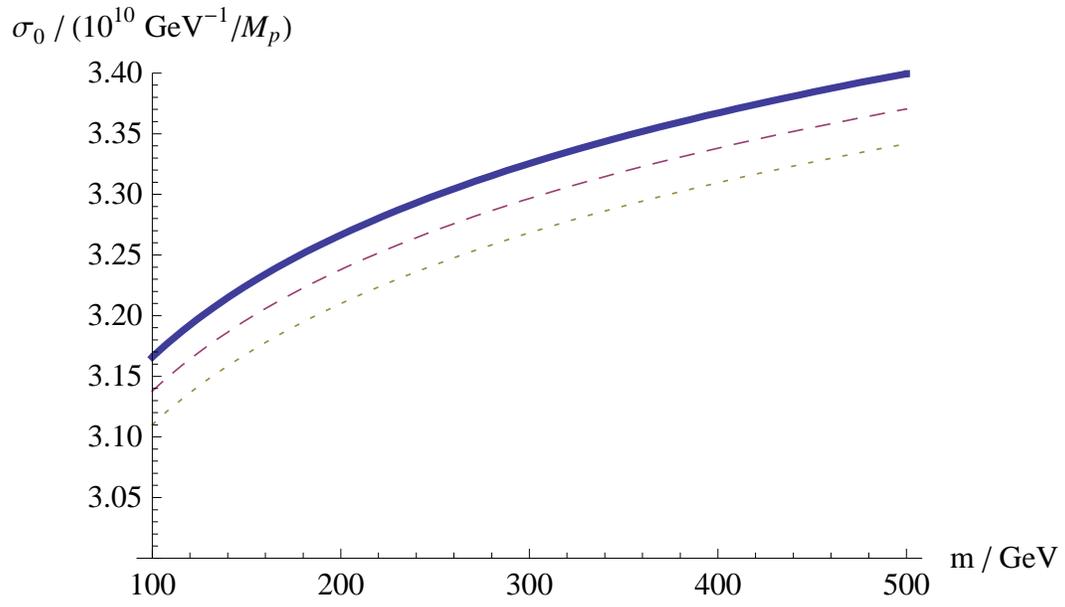,height=8cm,angle=0}}
\caption{Same as figure 3, but for CDM abundance $\Omega_{cdm} h^2=0.075$.}
\end{figure}

\end{document}